\documentclass[sigconf]{acmart}
\usepackage{tabularx}
\usepackage{graphicx}
\usepackage{subcaption} 
\AtBeginDocument{%
  }

\copyrightyear{2024}
\acmYear{2024}
\setcopyright{rightsretained}
\acmConference[WWW '24 Companion]{Companion Proceedings of the ACM Web
Conference 2024}{May 13--17, 2024}{Singapore, Singapore}
\acmBooktitle{Companion Proceedings of the ACM Web Conference 2024 (WWW
'24 Companion), May 13--17, 2024, Singapore,
Singapore}\acmDOI{10.1145/3589335.3651245}
\acmISBN{979-8-4007-0172-6/24/05}

\acmSubmissionID{5124}

\begin{document}

\title{BoostER: Leveraging Large Language Models for Enhancing Entity Resolution}

\author{Huahang Li}
\orcid{0009-0000-3087-9608}
\affiliation{%
  \institution{Hong Kong Polytechnic University}
  \country{Hong Kong}
}
\email{hua-hang.li@connect.polyu.hk}

\author{Shuangyin Li}
\affiliation{%
  \institution{South China Normal University}
  \city{Guangzhou}
  \country{China}}
\email{shuangyinli@scnu.edu.cn}

\author{Fei Hao}
\affiliation{%
  \institution{Hong Kong Polytechnic University}
  \country{Hong Kong}
}
\email{ffaye.hao@polyu.edu.hk}

\author{Chen Jason Zhang}
\authornote{Chen Jason Zhang is the corresponding author.}
\affiliation{%
  \institution{Hong Kong Polytechnic University}
  \country{Hong Kong}
  }
 \email{jason-c.zhang@polyu.edu.hk}

\author{Yuanfeng Song}
\affiliation{%
  \institution{Webank Co. Ltd.}
  \city{Shenzhen}
  \country{China}
 }
\email{yfsong@webank.com}

\author{Lei Chen}
\affiliation{%
  \institution{Hong Kong University of Science and Technology}
  \country{Hong Kong}
  }
\email{leichen@cse.ust.hk}

\renewcommand{\shortauthors}{Huahang Li et al.}

\begin{abstract}
Entity resolution, which involves identifying and merging records that refer to the same real-world entity, is a crucial task in areas like Web data integration. This importance is underscored by the presence of numerous duplicated and multi-version data resources on the Web. However, achieving high-quality entity resolution typically demands significant effort. The advent of Large Language Models (LLMs) like GPT-4 has demonstrated advanced linguistic capabilities, which can be a new paradigm for this task. In this paper, we propose a demonstration system named \textbf{BoostER} that examines the possibility of leveraging LLMs in the entity resolution process, revealing advantages in both easy deployment and low cost. Our approach optimally selects a set of matching questions and poses them to LLMs for verification, then refines the distribution of entity resolution results with the response of LLMs. This offers promising prospects to achieve a high-quality entity resolution result for real-world applications, especially to individuals or small companies without the need for extensive model training or significant financial investment.
\end{abstract}




\maketitle

\vspace{-1em}
\section{Introduction}
Living in the ear of the Web, we are surrounded by a vast array of information. Yet, the Web information is often messy, duplicated, or represented in myriad forms but refers to the same entity. Table~\ref{tabduplicates} exhibits a Web application acting as an online repository for specialists, such as LinkedIn, but contains duplicated records in its database. Entity resolution focuses on identifying and merging the duplicated or associated records and creating a complete and precise representation of each entity. This process can utilize a range of techniques, such as deterministic matching, which depends on exact attribute matches, and probabilistic or machine learning-based methods. These approaches take into account multiple aspects and utilize statistical models to calculate the matching probabilities \cite{winkler2014matching, elmagarmid2006duplicate}. A typical entity resolution workflow includes several crucial steps for ensuring accurate and reliable outcomes, which involve data preprocessing, record blocking, pairwise comparison, scoring and thresholding, and eventually clustering \cite{christendata}. The outcome of entity resolution is a refined database possessing consolidated and unique representations of distinct real entities.  
This standard workflow is the basis for entity resolution tasks and has been commonly deployed in diverse areas to address the challenge of discovering and combining replicated data records.

However, most of the previous efforts focused on constructing entity resolution tools have approached the challenge as a classification problem. Within this framework, classifiers are carefully constructed to efficiently and precisely distinguish between pairs that are duplicates and those that are distinct, thereby categorizing these pairs based on similarity \cite{kdd2003}. As a result, achieving optimal performance typically requires the design and training of specific models tailored to particular datasets. This approach restricts the model's applicability to limited scenarios and presents difficulties in adapting them to different domains. Though there are initiatives to develop a more generalized entity resolution model, the performance of these broader models remains imperfect \cite{tang2022generic}. 

A practicable method for enhancing entity resolution results involves integration with external knowledge sources, such as verification by crowd workers or insights from LLMs \cite{wang2012crowder, narayan2022can}. In recent years, we have witnessed significant advancements in LLMs like GPT-4, Claude2, etc.. These LLMs, trained on vast and diverse datasets, excel at capturing intricate linguistic patterns, contextual nuances, and semantic meanings \cite{zhao2023survey, jiang2023probabilistic}. An impressive strength of LLMs is their capability in contextual understanding and disambiguation, which have proven useful in handling ambiguous references and inconsistencies in entity attributes—elements that have traditionally been challenging in this domain. This endows them with a remarkable capability to match and compare entity attributes across varied records, enabling more precise and thorough outcomes. However, most LLMs, such as OpenAI's GPT-4, are charged for online API requests based on the total number of tokens in both the input (prompt or question) and the output (model's response)\footnote{https://openai.com/pricing}. For example, if an API request includes an input of 10 tokens and yields an output of 20 tokens, the billing will be for 30 tokens in total. The token, which is the smallest unit processed by the model, can vary in size, representing anything from individual words to subwords or characters. Given that posing all matching questions could result in substantial costs, efficiently and effectively leveraging LLMs for enhancing entity resolution has turn into a crucial challenge.

\begin{table}[t]
\caption{Database Contents: Profiles of Professionals in the Web Application. Ground Truth: Records $r_1$ and $r_2$ correspond to the same individual, as do records $r_3$ and $r_4$.}
\label{tabduplicates}
\centering
\begin{tabular}{|c|p{1.2cm}|p{1.5cm}|p{1.2cm}|p{1.3cm}|p{0.8cm}|}
\hline
\textbf{ID} & \textbf{Name} & \textbf{Email} & \textbf{Title} & \textbf{Company} & \textbf{Loc. } \\
\hline
$r_1$ & John Doe & johndoe @email.com & Software Engineer & TechCorp & SF\\
\hline
$r_2$ & J. Doe & johndoe @email.com & Software Engineer & TechCorp LLC & SF, CA \\
\hline
$r_3$ & Jane Smith & janesmith @email.com & Project Manager & Innovate Tech & New York \\
\hline
$r_4$ & Jane S. & janesmith @email.com & PM & Innovate Tech & NY \\
\hline
$r_5$ & Johnathan Doe & johnathan.d @email.com & Developer & TechCorp & SF \\
\hline
\end{tabular}
\vspace{-10pt}
\end{table}

\begin{figure}[t]
  \centering
  \vspace{-1em}
  \includegraphics[width=5cm]{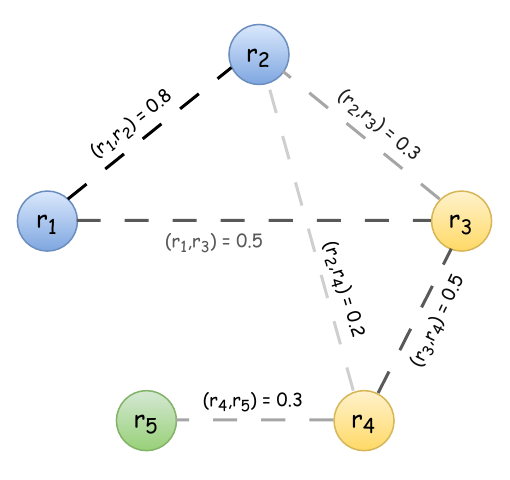}
  \vspace{-1em}
  \caption{An illustration of Possible Matches (linkages) in Table~\ref{tab:partitions}. The Probability of each linkage is the cumulative sum of its occurrences across Possible Partitions.}
  \label{example}
  \vspace{-2em}
\end{figure}

In this paper, we introduce a cost-effective demonstration system named \textbf{BoostER},  which makes full use of the power of LLMs as a service. In our theoretical approach, every possible partition of the entity resolution results is taken into account to ensure the result set encompasses any conceivable scenario. Each possible partition is assigned with a probability that reveals the likelihood of it being accurate, as shown in Table~\ref{tab:partitions}. In practical applications, partitions are generated by basic entity resolution tools. We aggregate all the entity resolution results and then normalize these probabilities to sum up to 1. Subsequently, the probability of each potential matching pair is calculated by summing up the probabilities across its possible partitions. As depicted in Figure~\ref{example}, each record is denoted by a node in a graph, and each link associated with a probability is represented as the potential matching pair. Therefore, the objective of entity resolution can be viewed as identifying potential linkages between these nodes. We adopt Shannon entropy to gauge the uncertainty of possible partitions. The underlying principle is that reducing entropy in a fixed system requires an external energy source, which is precisely the role LLMs fulfill. Our tailored greedy algorithm selects an optimal set of matching questions within the given budget, balancing the effectiveness of the matching questions against the cost of the number of tokens. Subsequently, we request these matching questions to LLMs for verification. Based on the responses from LLMs, we refine the probability distribution of possible partitions and recalculate the probabilities of possible matches. After several iterations of adjustments or upon exhausting the budget, a precise distribution of the entity resolution results is achieved.

\begin{table}[t]
    \centering
    \caption{Probability Distribution of Possible Partitions in Table~\ref{tabduplicates}. The sum of all these probabilities equals to 1.}
    \label{tab:partitions}
    \begin{tabularx}{0.85\columnwidth}{Xlc}
        \hline
        Possible Partition & Probability \\
        \hline
        \(P_1 = \{(r_1,r_2), (r_3,r_4), (r_5)\}\) & 0.5 \\
        \(P_2 = \{(r_1,r_2,r_3), (r_4,r_5)\}\) & 0.3 \\
        \(P_3 = \{(r_1,r_3), (r_2,r_4), (r_5)\}\) & 0.2 \\
        \hline
    \end{tabularx}
\end{table}

\begin{figure*}[t]
  \centering\includegraphics[width=0.9\textwidth]{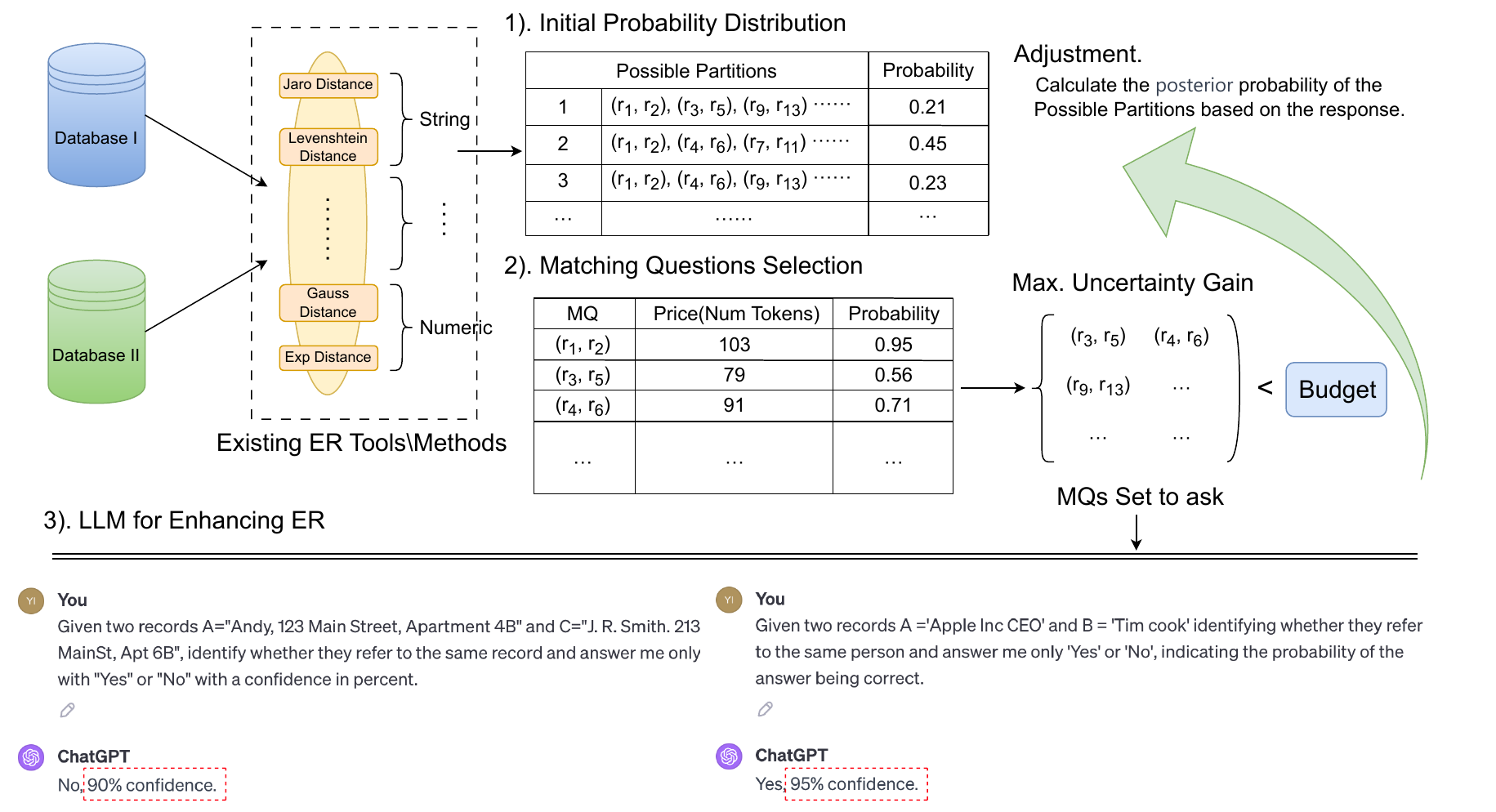}
  \caption{The Workflow of BoostER.}
  \label{LLM enhanced ER}
  \vspace{-10pt}
\end{figure*}

\section{The BoostER Framework}
In this section, we present a detailed description of the BoostER framework. As depicted in Figure~\ref{LLM enhanced ER}, BoostER encompasses several key steps: (1) Initialization of the Probability Distribution, (2) Selection of an Optimal Set of Matching Questions, (3) Verification by LLMs and Refinement of the Probability Distribution based on the responses. The technical specifics of each step are elaborated in the subsequent parts of this section.

\subsection{Probability Distribution Initialization}

The BoostER workflow begins by inputting records with duplicates, which may originate from multiple databases. Upon receiving these records, BoostER initially employs existing entity resolution tools to create a set of possible matches, each accompanied by a corresponding probability. An appropriate threshold is resorted to filter pairs with low probabilities. Then, to initiate the probability distribution of the possible partitions, we treat each pair as independent of the others, regarded as the Bernoulli distribution. Subsequently, we utilize Shannon entropy to assess the uncertainty of the results. A higher entropy value indicates greater incongruity among various entity resolution tools, signifying high uncertainty of the results.

\subsection{Matching Questions Selection}

We quantified the occurrence of matching pairs across all possible partitions to determine the probability of each pair. This probability, derived from statistical analysis, indicates the likelihood of a potential linkage between two records. A probability closer to 1 or 0 suggests higher certainty about the matching pair, whereas a value near 0.5 indicates greater uncertainty. Intuitively, one might select questions with the most uncertainty, for example, those with probabilities near 0.5. However, the selection process is complicated by the correlation between matching pairs. For instance, if we already get ``$r_1 = r_2$" and ``$r_2 = r_3$",  then querying whether ``$r_1 = r_3$" is unnecessary due to the transitivity property.

To find an optimal set of matching questions that leads to the most uncertainty reduction of possible distribution, we have established that the reduction in uncertainty is equivalent to the entropy of the answer set of these questions, denoted as $D_A$, and irrespective of the answering capability of LLMs. For a detailed proof, please refer to \cite{li2024leveraging}. Thus, our problem turns out to be maximizing the joint entropy of the answer set within a given budget. Considering the cost associated with each question, we have devised a price function to convert the question to a constant price based on OpenAI's Tokenizer. Following these steps, we can effectively calculate both the anticipated benefits and the associated costs for any given set of matching questions.

The selection problem can be easily proven to be NP-hard. However, given that the joint entropy is a sub-modular function, our tailored greedy algorithm is capable of achieving near-optimal results. Specifically, it guarantees an approximation ratio of $(1-1/e)$. This constitutes the main contribution of our work.

\begin{figure*}[t]
\subfloat[(I): Input the data and select the appropriate tools.]{
  \includegraphics[width=0.45\textwidth]{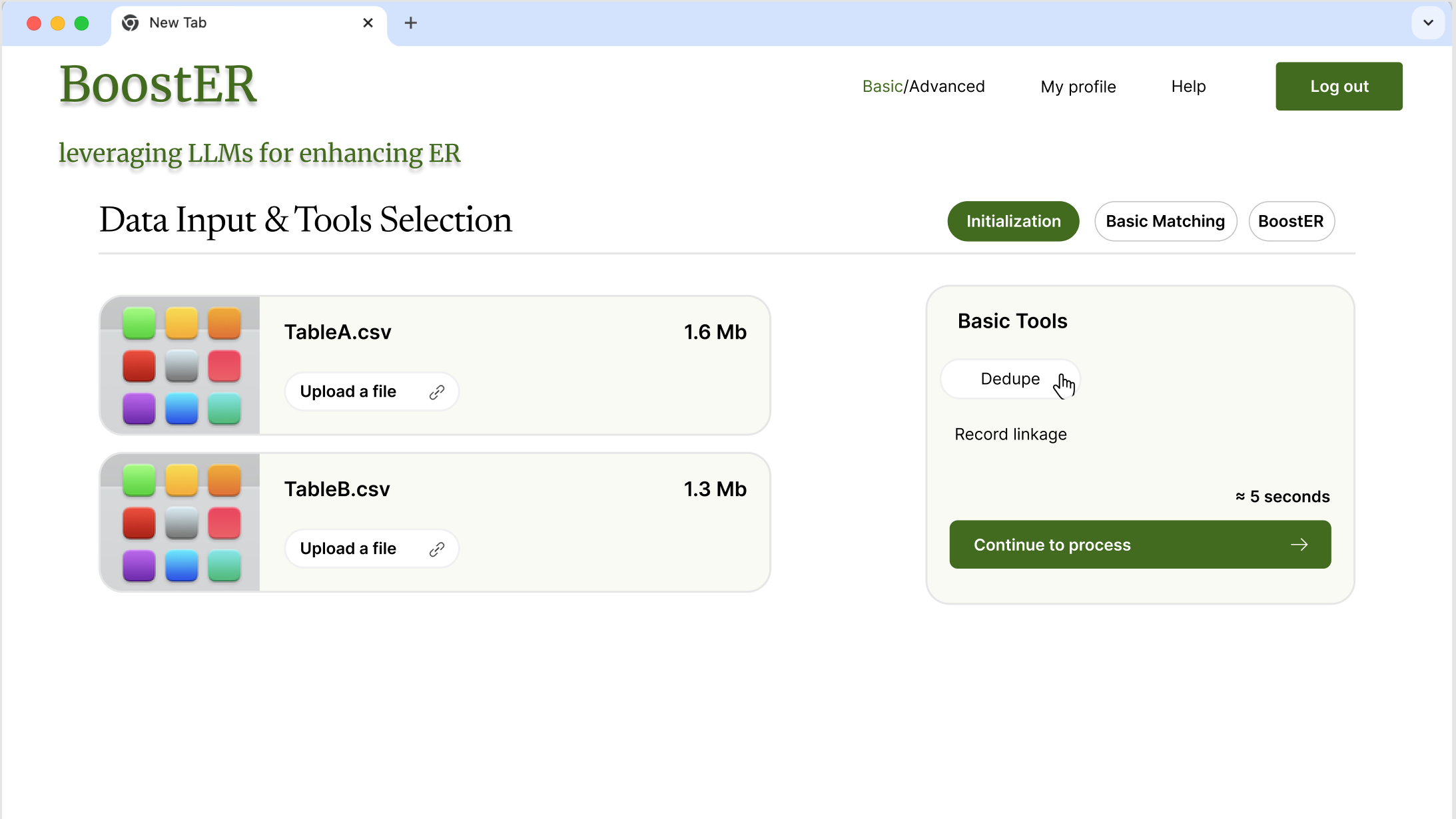}
  }
\hspace{10mm}
\subfloat[(II): Choose the LLM and configure your budget. ]{
  \includegraphics[width=0.45\textwidth]{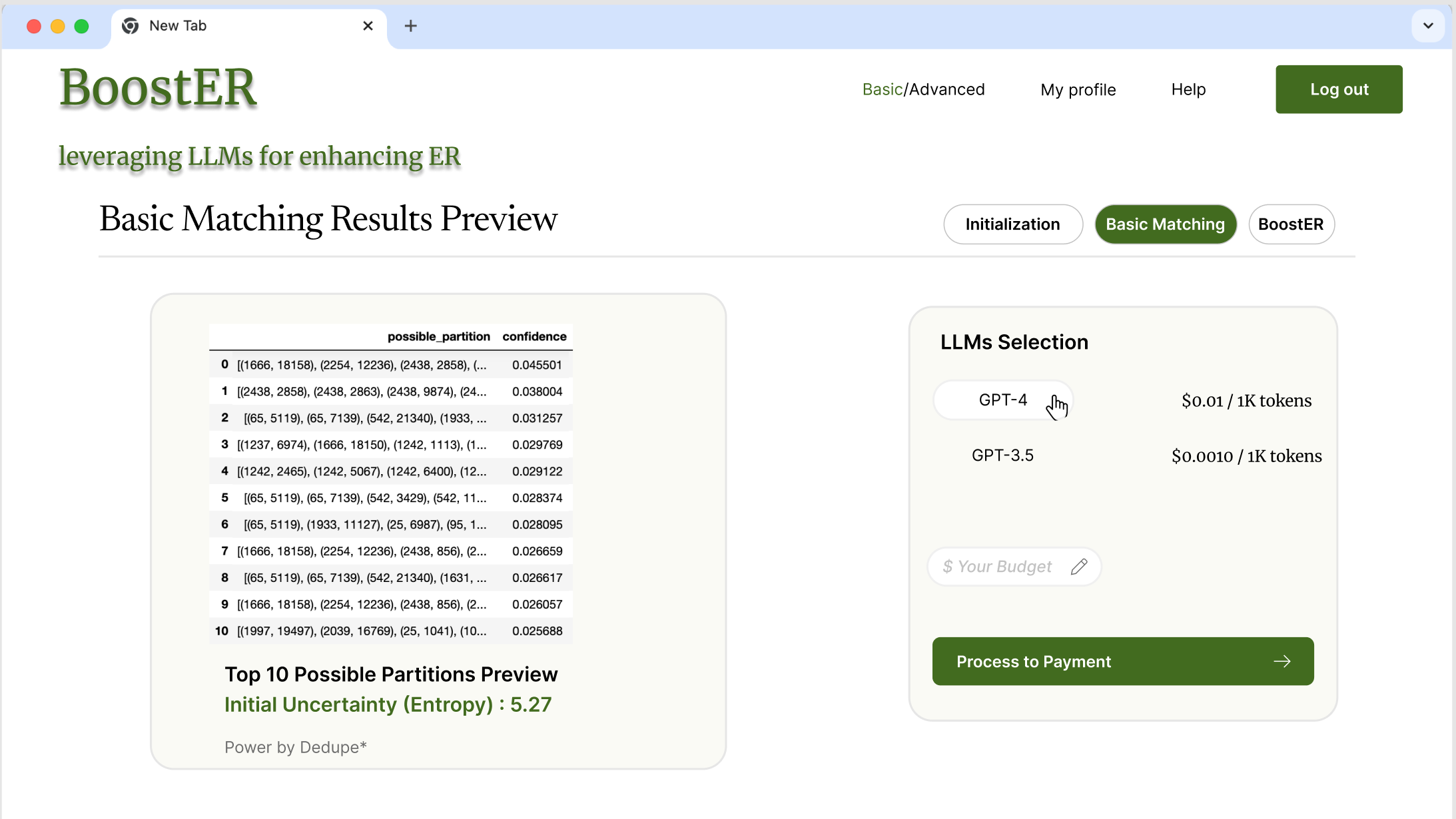}
}
\vspace{-1em}
\caption{BoostER Demo}
\label{demo}
\end{figure*}

\begin{figure}[t]
\vspace{-1em}
\subfloat[(III): Observe and analyze the results.]{
    \includegraphics[width=0.45\textwidth]{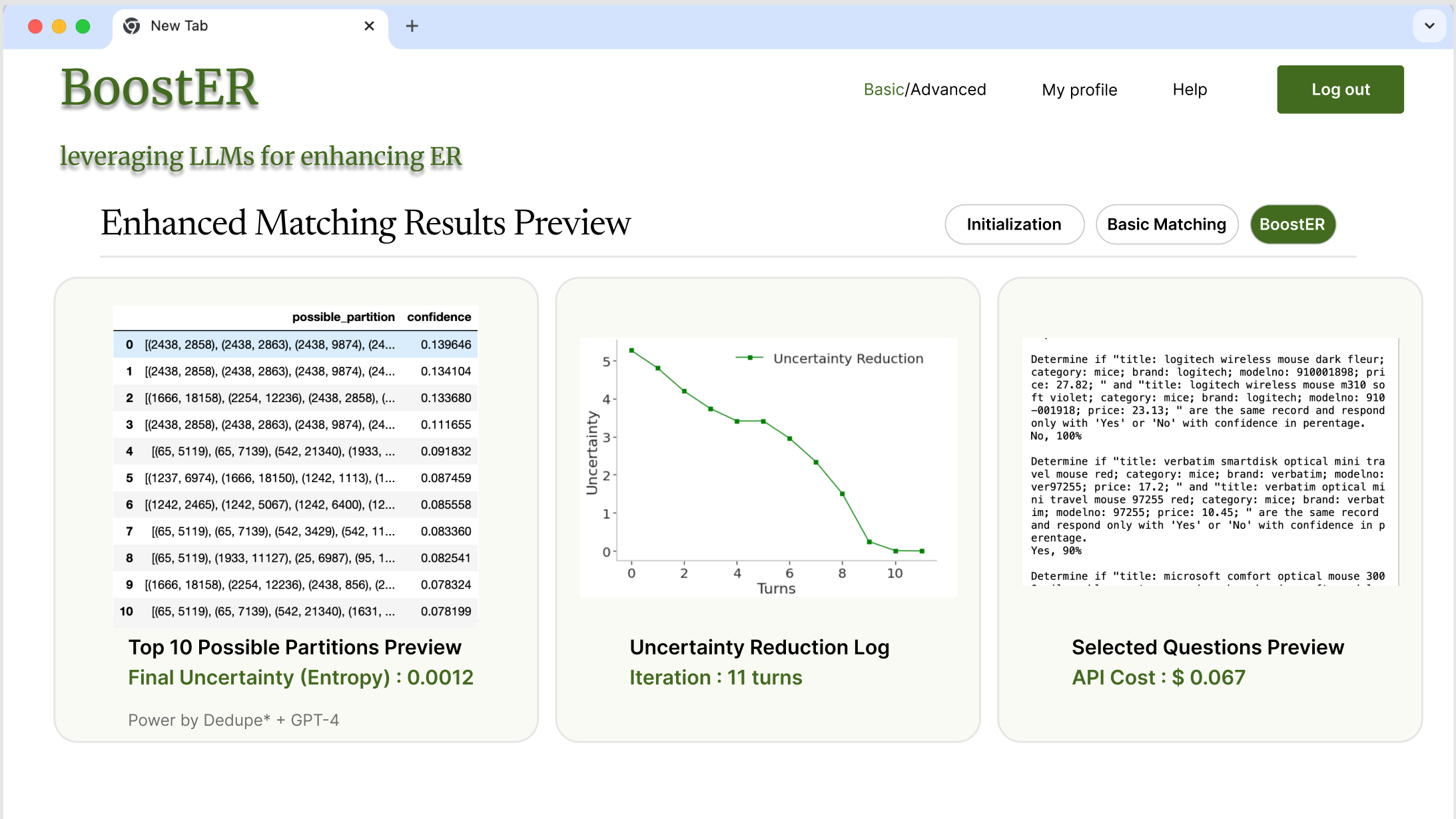}
}
\vspace{-1em}
\end{figure}

\vspace{-10pt}
\subsection{Adjustment with LLMs Response}
In real applications, even with the most advanced LLM, errors can still occur. The capability of an LLM, denoted as $\Theta$, can be defined by the expected accuracy rate from the answers generated in specific tasks. This can be estimated by performing sample questions before starting. Our BoostER framework is designed to be error-tolerant. Below, we provide a running example demonstrating the adjustment process with LLMs response.

\textbf{Running Example}: Proceeding with the example in Figure~\ref{example}, assume the record pair $(r_3,r_4)$ is recognized as correct by the LLM and the LLM's capability is 90$\%$, we can deduce:
\begin{equation}
\vspace{0.5em}
\begin{split}
&\mathcal{P}(P_1|Ans:(r_3,r_4) \ is \  answered \ from \  LLM)\\ =&\frac{\mathcal{P}(P_1)\mathcal{P}(Ans|P_1)}{\mathcal{P}(Ans)}\\
=& \frac{\mathcal{P}(P_1)\mathcal{P}(\Theta)}{\mathcal{P}(r_3,r_4)\mathcal{P}(\Theta) + (1- \mathcal{P}(r_3,r_4)\mathcal{P}(1-\Theta)} \\
= &\frac{0.5 * 0.9}{0.5*0.9+0.5*0.1} = 0.9,
\end{split}
\end{equation}
where $\mathcal{P}(\cdot)$ is the probability function. Similarly, for $P_2$, we have 
\begin{equation}
\vspace{0.5em}
\begin{split}
&\mathcal{P}(P_2|Ans:(r_3,r_4) \  is \  answered \ from \  LLM) )\\ =& \frac{\mathcal{P}(P_2)\mathcal{P}(Ans|P_2)}{\mathcal{P}(Ans)}\\
=& \frac{\mathcal{P}(P_2)\mathcal{P}(\Theta)}{\mathcal{P}(r_3,r_4)\mathcal{P}(1-\Theta) + (1- \mathcal{P}(r_3,r_4))\mathcal{P}(\Theta)} \\
= &\frac{0.3 * 0.1}{0.5*0.1+0.5*0.9} = 0.06.
\end{split}
\end{equation}

$\mathcal{P}(P_3)$ can also be obtained as the above and the final result is $\mathcal{P}(P_3) = 0.04$. As a result, the uncertainty of the possible partitions is significantly reduced from 0.464 $\rightarrow$ 0.186. Since $\Theta = 90\%$, this reduction in uncertainty is slightly less than what would be achieved with error-free responses. However, this demonstrates that even imperfect answers can substantially aid in diminishing uncertainty. With repeated iterations of this process, when either the entropy reduction ceases or the budget is exhausted, the more dependable distribution of possible partitions is acquired. 

\section{Demonstration}
This demonstration aims to provide an interactive system for users to enhance entity resolution results by leveraging LLMs through the techniques described previously. The main three steps are depicted in Figure~\ref{demo}, and we will introduce them respectively.

\vspace{0.5em}
\noindent
\textbf{Initialization:} BoostER supports multiple data resources in .csv format, allowing for easy uploads via the ``Upload" button. Currently, the system integrates built-in tools such as Dedupe \cite{dedupe} and Record Linkage \cite{de_bruin_j_2019_3559043}. Users can select a specific tool and initiate the process by clicking the corresponding button. Subsequently, the program automatically executes multiple iterations with various parameters.

\vspace{0.5em}
\noindent
\textbf{Basic Matching:} This step presents the entity resolution results produced by the fundamental tools, showcasing the top 10 possible partitions in a table format. The Initialization is designed for basic users, this implementation allows for an introduction to the system's capabilities. In the advanced version, users have the option to upload their entity resolution results. They can then select a specific LLM and establish a budget for its use.

\vspace{0.5em}
\noindent
\textbf{BoostER:} In the final step, the improved entity resolution results are displayed, along with the operational logs of the BoostER system. These include the uncertainty reduction curve and a log of the selected questions.

\section{Conclusion}
In this paper, we design the BoostER framework, which provides a cost-effective application of LLMs in enhancing entity resolution results. The BoostER achieves maximal effectiveness within the given budget of API request. Our target users are small companies or individual users without the need for extensive model training or significant financial investment, but can obtain high-quality results with general tools and a small amount of money. In other words, we reduce the cost of obtaining entity resolution, both in terms of ease of use and cost of use. In future work, we will explore more prompting techniques to further improve the performance.

\begin{acks}
This work was partially supported by Major Program of National Language Commission  (WT145-39) and Natural Science Foundation of Guangdong (2023A1515012073). And this work was also supported from the following funding sources: PolyU (UGC) - P0045695, Innovation and Technology Fund (P0043294), PolyU-MinshangCT Generative AI Laboratory (P0046453), Research Matching Grant Scheme (P0048191, P0048183), and PolyU Start-up Fund (P0046703).
\end{acks}

\bibliographystyle{ACM-Reference-Format}
\bibliography{sample}

\end{document}